\documentclass[a4paper]{article}
\usepackage{iwslt18,amssymb,amsmath,epsfig}
\usepackage{subfig}
\usepackage{soul}
\usepackage{color}

\def\reg{{\rm\ooalign{\hfil
     \raise.07ex\hbox{\scriptsize R}\hfil\crcr\mathhexbox20D}}}

\title{Fine-tuning on Clean Data for End-to-End Speech Translation: \\  FBK @ IWSLT 2018}

\makeatletter
\def\name#1{\gdef\@name{#1\\}}
\makeatother
\name{{\em Mattia Antonino Di Gangi$^{1,2}$, Roberto Dess\`i $^1$*\thanks{* Work performed during an internship at FBK}, Roldano Cattoni$^2$,}\\
      {\em Matteo Negri$^2$, Marco Turchi$^2$}}

\address{$^1$University of Trento, Italy \\ $^2$Fondazione Bruno Kessler, Italy\\
{\small \tt $^1$roberto.dessi@unitn.it} \\
{\small \tt $^2$\{digangi,cattoni,negri,turchi\}@fbk.eu} 
}
\begin{document}
\maketitle
%

\begin{abstract}
This paper describes FBK's submission to the end-to-end English-German speech translation task at IWSLT 2018. Our system relies on a state-of-the-art model based on LSTMs and CNNs, where the CNNs are used to reduce the temporal dimension of the audio input, which is in general much higher than machine translation input.
Our model was trained only on the audio-to-text parallel data released for the task, and fine-tuned on cleaned subsets of the original  training corpus. The addition of weight normalization and label smoothing improved the baseline system by $1.0$ BLEU point on our validation set. The final submission also featured checkpoint averaging within a training run and ensemble decoding of models trained during multiple runs. On test data, our best single model obtained a BLEU score of $9.7$, while the ensemble obtained a BLEU score of $10.24$. 
\end{abstract}

\section{Introduction}
End-to-end speech translation (that is, the direct translation of an audio signal without intermediate transcription steps) 
has recently gained increasing interest in the scientific community thanks to the recent advances of neural approaches in the related ASR and MT fields
\cite{berard2016listen,berard2018end,anastasopoulos2018tied,weiss2017sequence,anastasopoulos2018leveraging}. Effective approaches to the task can become a useful solution to deal with languages that do not have a formal writing system~\cite{duong2016attentional}, as it is possible to create a collection of spoken utterances with their respective translations in a more common language. We can also expect that, in the future, end-to-end speech translation systems will overcome problems related to the cumulative effect of  speech recognition errors introduced in pipelined architectures.
%
FBK's submission to the IWSLT 2018 Speech Translation  (ST) task
relies on a single model that takes as input features extracted from an English audio signal and returns as output a written translation in German. As the input is not in raw wave form, one might argue that the ``end-to-end'' denomination does not fit in this formulation of the task. Nevertheless, since feeding the network with  the input features released by the task organizers was allowed, we adhere to the looser definition of ``end-to-end'' implicit in this year's task formulation.\footnote{Our work has been pursued during a summer project with the goal of gaining hands-on expertise in this new promising field with the simplification of a standardized data set.} 

Our system was trained using the state-of-the-art sequence-to-sequence model based on LSTMs and CNNs introduced in~\cite{berard2018end}. 
%
%
Considering the high number of experiments to run, and the high number of epochs needed to train a speech translation model (up to $87$ in the case of our final submission), 
the model
was implemented using the \textit{fairseq}\footnote{http://github.com/facebookresearch/fairseq}~\cite{gehring2017convolutional} sequence-to-sequence learning toolkit from Facebook AI Research. The tool, which is tailored to NMT, was adapted to the ST task  
showing considerable reductions in training time compared to the same models implemented on other platforms (from hours to minutes in the processing of the same amount of training instances).

One of the main challenges we faced was how to maximize the usefulness of the available training data by weeding out noisy (and potentially harmful) instances. For this purpose, we developed the two data cleaning procedures  described in Section \ref{sec:preproc}. The architectural choices and the main implementation details of our system are described in Section~\ref{sec:model}.
In Section \ref{sec:exp}, we report the results on our validation set, which were obtained by using different data conditions and hyper-parameters. 
Section \ref{sec:concl} concludes the paper with final remarks.





\section{Data Cleaning}
\label{sec:preproc}
Our submission was obtained by a model solely trained with the data released for the speech translation task.
Before building the model, we devoted particular attention to the quality of the training material, aiming to reduce the possible impact that noise in the data can have on training time and model convergence.  Indeed, the initial training set of $171,121$ instances comprised elements featuring either a partial alignment between the audio signal and the corresponding  transcription, or a skewed ratio between the number of feature frames and the characters in the transcription. To identify and weed out such noisy and potentially harmful training items,  we applied two cleaning procedures. Both the procedures take advantage of the available English transcriptions of the audio signals\footnote{Note that data cleaning is the only phase in which we used the English transcriptions. Being this step independent from the actual system training, our approach is still fully end to end.} and were  run in cascade, after the removal of $1,000$ items to be used as our development set.
As discussed in Section \ref{subsec:datasel}, though smaller in size, the resulting subsets of the original training corpus yielded performance improvements on development data, especially when used for fine-tuning a model trained on the original unfiltered corpus.

\subsection{Cleaning Based on Alignment}
Starting from the initial 
training corpus 
of $170,121$ instances (called ``Parallel'' henceforth), the first cleaning step was aimed to identify and remove the 
items featuring a poor alignment between the audio signal and the text. Assuming that the English and German texts are parallel, the potential noise introduced by such instances is represented by wrong transcriptions/translations (either totally inadequate or containing spurious words) of the original source signal. To identify them, our approach was to align each audio signal with the corresponding English transcription and then decide what to retain based on the alignment quality (i.e. considering unaligned words  as evidence of noise).  We performed the alignment on a sentence-by-sentence basis using Gentle,\footnote{https://lowerquality.com/gentle/} a forced aligner based on Kaldi.\footnote{http://kaldi-asr.org/index.html} After the alignment, we removed all the training instances in which at least one word in the transcription was not aligned with the corresponding audio segment. 
This strict cleaning policy (due to time limitations, we did not experiment with less aggressive strategies) resulted in the removal of $24,240$  instances, which
reduced 
the initial ``Parallel'' corpus
to $145,881$ items. Henceforth, the corpus resulting from this first cleaning step will be referred to as ``Clean 1''.

\subsection{Cleaning Based on Frames/Characters Ratio}

\begin{figure}
\centering
\includegraphics[width=0.46\textwidth,height=5.5cm]{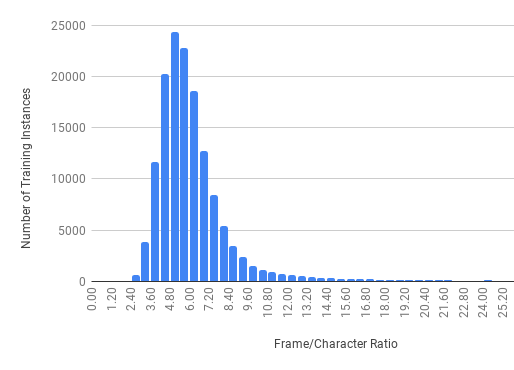}
\caption{Distribution of the training instances in terms of the ratio between the number of feature frames and the characters in the transcription.}
\label{fig:ratioplot}
\end{figure}

The second cleaning step was aimed to identify and remove from ``Clean 1'' the training instances featuring a skewed ratio between the number of feature frames and the characters in the transcription. In this case, the potential noise is due to portions of the original speech that 
correspond to long silences, background noise (e.g. laughter and applause), or words that are not present in the transcription/translation. 
To identify such possible outliers, looking at the ratios reported in Figure \ref{fig:ratioplot},  we decided to  cut the distribution  so to retain only the training instances belonging to ratio bins that contain at least $5,000$ items. The corresponding cutting values of $3.5$ and $7.5$ resulted in the removal of $29,898$ instances, which further reduced our training corpus 
to $115,983$ items.
Henceforth, the corpus resulting from our second cleaning step will be referred to as ``Clean 2''.

\section{Seq2seq Speech Translation model}
\label{sec:model}

\begin{figure}[t]
\centering
\includegraphics[width=0.5\textwidth,height=7cm]{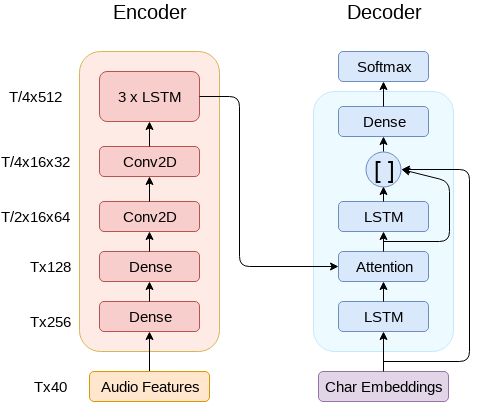}
\caption{Schema of our end-to-end model architecture. The numbers on the left represent the dimensionality of each encoder layer's output. The batch size is not written.}
\label{fig:schema}
\end{figure}

We re-implemented the seq2seq ST model introduced in~\cite{berard2018end}, which uses an encoder-decoder-attention architecture based mainly on LSTMs~\cite{hochreiter1997long}. The source-side input length is some order of magnitudes higher than the decoder side, and thus some reduction in the temporal dimension 
was performed using 2-D CNNs with stride (2, 2). The decoder is inspired by the early deep-transition decoder used in Nematus~\cite{sennrich2017nematus}, which stacks two LSTM units in a way that the single LSTMs are not recurrent by themselves, while the stack of 
the 
two is globally recurrent.
A schema of the model is depicted in Figure~\ref{fig:schema}.

\subsection{Encoder}
The input to the encoder is a variable-length audio sequence with $40$ features for each time step. At first, the input sequence is processed by two time-distributed densely-connected layers with size of $256$ and $128$ respectively,  each followed by a \textit{tanh} activation. 
The output of the densely-connected layers is then processed by two stacked 2-dimensional convolutional layers, each having a $3\times 3$ kernel and stride $= 2$. Let $n$ be the sequence length and $f$ be the number of input features to the first convolutional layer. The output of the first convolution is of size $(16, n/2, f/2)$ and for the second convolution is of size $(16, n/4, f/4)$. The $16$ filters are then flattened to obtain an output of size $(n/4, 4\times f)$, which is 
subsequently 
processed by a stack of three bidirectional LSTM layers~\cite{schuster1997bidirectional}. The initial state of the LSTM is initialized as a zero vector at the beginning of the training, but then it is optimized via back-propagation together with the rest of the network. We found that training the initial state gives a boost in performance and speeds up the model convergence.

\subsection{Decoder}

The decoder consists of a two-layered deep-transition LSTM~\cite{sennrich2017nematus} followed by a deep output layer~\cite{pascanu2013difficulty}.
The input of the first layer is the character embedding of the last character. The output of the first layer is used as a query vector to compute an attention over the last layer of the encoder. The attention output is then used as input to the second LSTM layer. The hidden and cell states received as input by the two LSTM layers are, for every time step, the last hidden and cell states produced by the other LSTM layer. 
The last encoder output is averaged over the time dimension and this new tensor is passed as input to two different densely-connected layers with \textit{tanh} non-linearity. The two functions compute the initialization of the hidden and cell states for the first LSTM layer.
The deep output is a densely-connected nonlinear function, which takes as input the concatenation of the LSTM output, the attention output and the current symbol (character) embedding, and outputs a tensor of size $512$. This tensor is finally multiplied by a second character embedding matrix to compute the scores over the whole vocabulary.

\subsection{Attention}

The attention layer computes a distribution of weights that sums up to $1$ for the encoder output sequence (soft attention) with no positional information (global attention). The scores for each encoder position are computed according to their relevance with respect to the decoder state. The relevance score is computed using the general attention score proposed in~\cite{luong2015effective}.

\subsection{Increased Regularization}
Due to the small size of the training data, we found useful to 
apply 
some regularization tricks. The first and more common technique is the dropout applied to 
each 
layer \cite{gal2016theoretically}. 
Instead of  variational dropout \cite{kingma2015variational}, we preferred to use the fastest implementation of LSTMs provided by the Pytorch library, which uses regular dropout. 

Besides dropout, we applied weight normalization and label smoothing as additional techniques 
for
regularization. 
Weight normalization~\cite{salimans2016weight} is a simple technique that decomposes the parameter matrices into their magnitude and direction components in order to easily produce a transformation that scales the weights and reduces the gradient covariance to zero. The result is a faster convergence and a limitation of the weight space, which has a regularizing effect. 

Label smoothing~\cite{szegedy2016rethinking} smooths the cross-entropy cost function by 
giving a weight of $0.9$ to the probability of the correct symbol, and $0.1$ to the sum of the probabilities of the other symbols. Label smoothing makes the model less confident on its predictions, producing a regularizing effect.
In NMT, it has been observed that, despite the increased loss and perplexity usually obtained with this technique, the translations are usually better~\cite{chen2018best} and end up in improved BLEU~\cite{papineni2002bleu} scores.

\section{Experiments}
\label{sec:exp}
In this section we summarize the experiments that motivated our choices for the final submission.
Since the goal of our participation was
to explore the potential of a single end-to-end model that can translate directly from audio signals,
we used as training data only the Speech Translation TED Corpus that was released for the task. No pre-training has been performed on different types of data (such a pre-training would in fact rely on ASR data). 
All our models were trained using the Adam optimizer \cite{kingma2014adam} with learning rate of $0.001$, and values for $\beta_1$ and $\beta_2$ of $0.9$ and $0.999$. We applied dropout of $0.2$ to all layers, including the input features. The norm of the gradients was clipped to $5$.
All the models have been trained until convergence according to the loss on
a held-out set of $1,000$ sentences (see Section~\ref{sec:preproc}).
The results achieved by each model on the validation set are reported Tables \ref{tab:result_ds}--4.

At first, we 
experimented with 
the reference implementation of the sequence-to-sequence model\footnote{https://github.com/eske/seq2seq} that is based on Tensorflow~\cite{abadi2016tensorflow}.
However, with about $3.5$  hours per epoch on a single NVIDIA GTX-1080 GPU, its training time resulted to be incompatible with the need of quickly testing a range of alternative solutions. 
%
To avoid this bottleneck, we re-implemented the same model within the \textit{fairseq} toolkit, which is highly optimized to significantly reduce training time. Our re-implementation was indeed faster, with a reduction of the training time to about $30$ minutes per epoch for the largest version of the training corpus (``Parallel''), and about $20$ minutes per epoch for the smallest one (``Clean 2''). The wall clock time of a single training run was around $30$ hours, with a maximum of $10$ additional hours for the fine-tuning.

\begin{table}[]
\centering
\begin{tabular}{l|l}
\textbf{Data} & \textbf{Val. BLEU} \\ \hline
Parallel      & 8.54               \\
Clean 1       & 8.98               \\
Clean 2       & 8.54              
\end{tabular}
\caption{Results of the base model over the three different versions of the dataset.}
\label{tab:result_ds}
\end{table}

\subsection{Dataset Selection}
\label{subsec:datasel}
In the first 
round
of experiments, we were interested in 
understanding the  impact of the data cleaning procedures described in Section \ref{sec:preproc}.
To this aim, we trained the base 
system 
on the three different versions of the dataset (i.e. ``Parallel'', ``Clean 1'' and ``Clean 2'') and evaluated 
the resulting models 
on the same validation set. The results listed in Table \ref{tab:result_ds} show that Clean 1 provides us with the best result, but Clean 2 leads to a result equivalent to Parallel despite using about $36\%$ less data. Thus, we decided to use Clean 2 for the following experiments in order to have faster training cycles.

\begin{table}[t]
\centering
\subfloat[Dataset fine-tuning]{
\begin{tabular}{l|l}
\textbf{Data}                       & \textbf{Val. BLEU} \\\hline
P $\rightarrow$ C1                  & 9.55               \\
P $\rightarrow$ C2                  & 9.85               \\
C1 $\rightarrow$ C2                 & 9.89               \\
P $\rightarrow$ C1 $\rightarrow$ C2 & \textbf{10.14}             
\end{tabular}
\label{tab:results_finetuning}}\\
\subfloat[Restart strategy]{
\begin{tabular}{l|l}
\textbf{Strategy} & \textbf{Val. BLEU} \\\hline
Adam annealing    & 9.11               \\
NAG annealing     & 8.74              
\end{tabular}
\label{tab:results_restart}}
\caption{(a) Results for the base model in different fine-tuning conditions. P stands for Parallel, C1 for Clean 1 and C2 for Clean 2. Only the last row refers to a double step of fine-tuning. (b) Results with two different restart strategies for the model trained on Clean 2.}

\end{table}
\subsection{Dataset Fine-tuning and Restart Strategy}
In this subsection we 
address two questions. The first one is whether it is useful to fine-tune a model trained on a larger dataset 
by using  
a smaller and  cleaner 
subset of the same corpus.
The second question is whether a restart strategy with learning rate annealing can improve the performance. 

The first question 
was 
addressed by restarting the training of 
the 
model by using the new, smaller dataset as training set, but with the same training policy and hyper parameters. 
The results listed in Table \ref{tab:results_finetuning} show that  fine-tuning the model on cleaner data always helps. In particular,
fine-tuning on Clean 2 (which is smaller but of higher quality) always results in
better performance,
especially in the case of a double step of fine-tuning (P $\rightarrow$ C1 $\rightarrow$ C2).
Interestingly, also using only the clean data (C1 $\rightarrow$ C2) yields better results than training the initial model on the original Parallel corpus.

To address the second question, we used the model trained on Clean 2 and restarted the training on the same training set with a policy of learning rate annealing. To this aim, the learning rate was multiplied by $0.5$ every time  the validation loss did not improve over the best one computed so far~\cite{bahar2017empirical}. We experimented using both Adam with annealing and SGD with Nesterov Accelerated Gradient (NAG)~\cite{sutskever2013importance} with annealing. 
The results listed in table~\ref{tab:results_restart} show that, though Adam with annealing yields a better model, both the BLEU scores are at least $0.45$ points less than the worse model with fine-tuning.

\begin{table}[]
\begin{tabular}{l|l}
\textbf{Model}              & \textbf{Val. BLEU} \\ \hline
AST Seq2Seq                 & 8.54               \\
+ Weight Normalization (WN) & 8.69               \\
+ Label Smoothing (LS)      & 8.74               \\
+ Sigmoid Attention         & 8.44               \\
+ WN and LS                 & \textbf{9.69}              
\end{tabular}
\caption{Results on the data cleaned with two cleaning steps.}
\label{tab:results_feats}
\end{table}

\subsection{Features Exploration}
In this 
round 
of experiments we trained our base model on the Clean 2 dataset and compared its result with models that have weight normalization, label smoothing, sigmoidal attention instead of softmax attention, and weight normalization and label smoothing together. The results on the validation set, which are listed in Table~\ref{tab:results_feats}, show that both weight normalization and label smoothing  give a small contribution, while the sigmoidal attention slightly decreases the translation quality. Moreover, the joint addition of label smoothing and weight normalization gives a sensibly higher boost, suggesting that the models need high regularization. Considering the scarce amount of data, the need for high regularization was expected. However, it is interesting to note that by increasing the dropout to $0.3$ the base model converges to a much worse point.\footnote{Observed in preliminary experiments, not reported here.} From now on, we call the model with weight normalization and label smoothing ``full modell''.

\begin{table}[]
\centering
\subfloat[]{
\begin{tabular}{l|l}
\textbf{Data} & \textbf{BLEU} \\\hline
Parallel      & 4.66                   \\
Clean 1       & 9.69               \\
Clean 2       & 9.69              
\end{tabular}
\label{tab:results_wnls_data}}\\
\subfloat[]{
\begin{tabular}{l|l|l|l}
\textbf{Data}                       & \textbf{Best}  &  \textbf{Avg}  & \textbf{Test}\\\hline
P $\rightarrow$ C1                  & 10.26     &   10.46  & -   \\
C1 $\rightarrow$  C2                  & 9.71      &   10.42  & -    \\
P  $\rightarrow$  C2                  & 10.63     &   10.90  &  9.70    \\
P $\rightarrow$ C1 $\rightarrow$ C2 & 10.41     &   10.78  & -    \\
+ Adam annealing                    & 10.50     &   10.59  & -    \\\hline
Ensemble of 4                       &  -        &   \textbf{11.60}     &  \textbf{10.24} \\ 
\end{tabular}
\label{tab:results_wnls_finetune}}
\caption{Results using different versions of the dataset for training our model with weight normalization and label smoothing.}
\label{tab:results_wnls}
\end{table}
\subsection{Experiments with Full Model}
Once  we found that the full model is clearly better than the others, we replicated the experiments on all the datasets with the new model. In the second column of Table \ref{tab:results_wnls_data}, we can see that this model is more sensitive to noise. In fact, training it with the ``Parallel'' set leads to poor performance in translation ($4.66$ BLEU), but this lower translation quality was not expected by looking at only the training and validation losses. Nonetheless, the fine-tuning of this model on  cleaner data, whose results are listed in table~\ref{tab:results_wnls_finetune}, leads to improvements ranging from $0.57$ to $0.94$ BLEU points with respect to the models trained only on the clean data. 

Unfortunately, the score of $10.63$ of the best model (P$\rightarrow$C2) represents only a limited improvement when compared with the best model in the second column of Table~\ref{tab:results_finetuning} (P$\rightarrow$C1$\rightarrow$C2), which improved from $8.54$ of the base model to $10.14$. The fifth row of Table~\ref{tab:results_wnls_finetune} shows the results when the last fine-tuning is performed using Adam with annealing instead of Adam with a fixed learning rate. 
Based on these results, we 
submitted our single best model (P $\rightarrow$ C2 Avg) as our contrastive submission.

\subsection{Checkpoint Averaging and Ensemble Decoding}
\label{sec:ensemble}
Checkpoint averaging consists in computing the average of different checkpoints of the same training. 
In \cite{junczysneural}, it has been shown that, in neural machine translation, it leads to a better translation quality than using a single model. 
For 
each 
model, we computed the BLEU score on the validation set for the last $10$ checkpoints, and averaged the weights of all the models whose results are less than $0.5$ BLEU points worse than the best one.
The improvement can be observed by comparing the Best and Avg columns of Table~\ref{tab:results_wnls_finetune}. 

We also performed ensemble decoding of models trained in different runs. The ensemble involved all the Avg checkpoints listed in table~\ref{tab:results_wnls_finetune}, except for ``C1$\rightarrow$ C2'', which was trained using a different vocabulary.  The ensemble of the $4$ models obtained a result of $11.60$ BLEU on the validation set.

\subsection{Submitted Systems and Results}
Based on the outcomes of the above experiments on development data, 
we opted for submitting the following systems:
\begin{itemize}
    \item \textbf{Primary}: ensemble of 4 systems (Section~\ref{sec:ensemble}).
     \item \textbf{Contrastive}: Checkpoint averaging of P$\rightarrow$C2 (Table~\ref{tab:results_wnls_finetune}).
\end{itemize}
The  result of the primary system is $11.60$ BLEU score on our validation set and $10.24$ on the test set, whereas the contrastive system scored, respectively, $10.90$ and $9.70$ in the validation and the test set.

\section{Conclusions}
\label{sec:concl}


We described FBK's participation in the end-to-end speech translation task at IWSLT 2018.
We have shown that data cleaning is useful in reducing the training time by discarding a good portion of the training data, while not hurting translation quality. 
We have also observed that fine-tuning a model using a cleaner dataset can bring improvements up to $1.6$ BLEU points. 
Moreover, regularizing the model with normalization and label smoothing can produce an improvement of more than $1.0$ BLEU point with clean datasets, but the same model fails to converge to a good point using all the data.
In addition, using checkpoint averaging and ensemble decoding gave us another gain of $1.0$ BLEU point. The final score on this year's test set is of $9.70$ and $10.24$ BLEU respectively for our best single model and for the primary submission based on ensemble decoding.
In order to improve the competitiveness of this system, our next experiments will include ASR for pretraining the encoder~\cite{bansal2018pre} or for multi-task learning~\cite{weiss2017sequence}.

\section{Acknowledgements}
We gratefully acknowledge the support of NVIDIA Corporation with the donation of the GPUs used for this research.

\bibliography{iwslt2018}
\bibliographystyle{IEEEtran}
\end{document}